\documentclass[showpacs,aps,prb,reprint,twocolumn]{revtex4-1}

\usepackage{commath}
\usepackage{amsmath,bm}
\usepackage{graphicx}
\usepackage{hyperref}
\usepackage{txfonts}

\bibpunct{[}{]}{,}{n}{}{}

\draft 

\begin{document}


\title{A resolution of the problem of additional boundary conditions}

\author{Hai-Yao Deng}
\author{Egor A. Muljarov}
\affiliation{School of Physics and Astronomy, Cardiff University, 5 The Parade, Cardiff CF24 3AA, Wales, United Kingdom}

\begin{abstract}  
Maxwell's boundary conditions (MBCs) were long known insufficient to determine the optical responses of spatially dispersive medium. Supplementing MBCs with additional boundary conditions (ABCs) has become a normal yet controversial practice. Here the problem of ABCs is solved by analyzing some subtle aspects of a physical surface. A generic theory is presented for handling the interaction of light with the surfaces of an arbitrary medium and applied to study the traditional problem of exciton polaritons. We show that ABCs can always be adjusted to fit the theory but they can by no means be construed as intrinsic surface characteristics, which are instead captured by a \textit{surface response function} (SRF). Unlike any ABCs, a SRF describes essentially non-local boundary effects.  Methods for experimentally extracting the spatial profile of this function are proposed. 
\end{abstract}

\maketitle

\begin{figure}
\begin{center}
\includegraphics[width=0.45\textwidth]{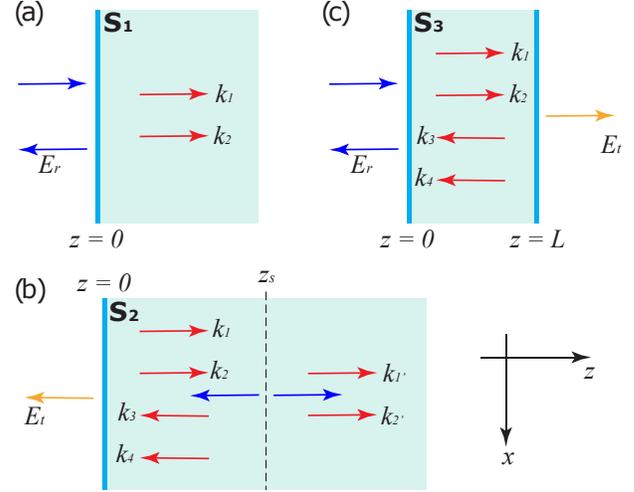}
\end{center}
\caption{Schematic of setups: (a) Setup S$_1$ where light incidents from vacuum upon an semi-infinite medium (SIM), (b) Setup S$_2$ where light is radiated from a source located at $z=z_s>0$ inside a SIM, and (c) Setup S$_3$ where light incidents from vacuum on a slab. The wave numbers $k_\alpha$ solve Eq.~(\ref{3}). \label{fig:0}}
\end{figure}

\textit{Introduction}. A light wave incident upon a dielectric gets partly reflected and partly transmitted. The textbook approach~\cite{landau,feynman} to determining the reflection and transmission amplitudes, denoted by $E_r$ and $E_t$, respectively, proceeds by writing down separately the expression for the waves in the vacuum and that in the dielectric, and then join them with Maxwell's boundary conditions (MBCs) at the surface. For instance, for a monochromatic beam normally incident on a medium [Fig.~\ref{fig:0} (a)] with frequency $\omega$ and wavenumber $k_0= \omega/c$, where $c$ the speed of light in vacuum, one may write, omitting the time dependence $e^{-i\omega t}$, for the electric field
$E(z<0) = e^{ik_0z}+E_re^{-ik_0z}$ and $E(z>0) = E_t e^{ikz}$. 
Here $k$ is the transmitted wavenumber, satisfying $\epsilon = (k/k_0)^2$, where $\epsilon$ is the dielectric constant. MBCs dictate the continuity of $E(z)$ and its derivative $E'(z)$, which determines $E_r$ and $E_t$. 

In 1957, Pekar claimed that MBCs were insufficient to determine the optical responses of a system of excitons~\cite{pekar1957}, for which $\epsilon$ is not a constant but a function of $k$. In the Lorentz oscillator model (LOM)~\cite{hopfield1958,stahl}, for example, one takes
\begin{equation}
\epsilon(k) \approx \epsilon_{b} + Q^2/(k^2-q^2). \label{2}
\end{equation}
Here $\epsilon_b$ denotes the background permittivity, $Q = \sqrt{2M\Delta/\hbar^2}$ and $q = \sqrt{2M(\omega+i\gamma - \omega_{ex})/\hbar}$, with $\Delta$, $\omega_{ex}$, $M$ and $\gamma$ being the exciton longitudinal-transverse splitting, transition energy, effective mass, and damping rate, respectively, and $\hbar$ is the reduced Planck constant. Waves propagating through such medium fulfill a dispersion relation given by
\begin{equation}
\epsilon(k) = \left(k/k_0\right)^2, \label{3}
\end{equation}
which admits two solutions, $k_1$ and $k_2$, representing waves propagating to the right and two other solutions, $k_3 = -k_1$ and $k_4 = -k_2$, for waves propagating to the left. Pekar hence wrote
$E(z>0) = \sum_{j = 1,2} E_{j} e^{ik_{j}z},$
where $E_{j}$ is the amplitude for the $j$-th transmitted wave, ending up with three unknowns, $E_r$, $E_1$ and $E_2$ but only two MBCs. He then introduced an additional boundary condition (ABC), imposing that the exciton polarization vanishes at the surface, in order to determine all the amplitudes.  
 
The practice of supplementing MBCs with ABCs has since been widely -- and sometimes unwittingly -- adopted for dealing with spatially dispersive media~\cite{moliner1979,forstmann1986}. Though popular, ABCs are unjustified and disputed~\cite{tignon2000,egor2002,deng2019,deng2020b,deng2017a,deng2017b}. Indeed, a whole zoo of ABCs, in addition to Pekar's original one, have been proposed in the literature~\cite{philbin2018}, while no \textit{a priori} criteria exists regarding which ABC should be selected for a given physical system. They are more of an experimental fitting machinery than a theoretical device. Efforts to remove ABCs without imposing extra approximations have failed so far~\cite{zeyher1972,chen1993,henneberger1998,nelson1999,zeyher1999,schmidt2016}.  

We contend that the current need of ABCs rests with an incomplete macroscopic view on physical boundaries such as surfaces and interfaces~\cite{agranovich1966,Johnson1976}.  
Here, we rectify this view and resolve the ABC problem. We derive a macroscopic theory for handling the optical responses of an arbitrary bounded medium without the need of ABCs. The theory is illustrated with the long-standing problem of exciton polaritons within the LOM -- extensions~\cite{victor1993,egor2002} are discussed in Ref.~[\onlinecite{si}] -- for three setups, S$_1$, S$_2$ and S$_3$, see Fig.~\ref{fig:0}. In both S$_1$ and S$_2$ a semi-infinite medium (SIM) with a single surface is considered. The difference is that, in S$_1$ the external source of light is placed outside the medium whereas in S$_2$ it is inside. In S$_3$ a slab medium is considered with light incident from outside. The physical effects of a surface are shown totally contained in a \textit{surface response function} $\mathcal{R}(z)$, which gauges the response of the surface to incoming polarization waves generated at a distance $z$ from the surface. We show that for each setup a set of ABCs can be derived to fit the theory, but no single set applies to all setups even with the same material. In addition, S$_1$ is shown insensitive to the full profile of $\mathcal{R}(z)$ but only probes some average, whereas S$_2$ and S$_3$ detect $\mathcal{R}(z)$ in full and can be used for experimentally extracting it. 
     
\textit{Macroscopic limit of a physical surface}. For clarity, let us consider a SIM bounded by a single surface lying in the $x$-$y$ plane. Microscopically, the system (i.e.~the medium plus the vacuum) naturally divides into three regions: the vacuum $z<0$, the surface region $0\leq z<d_s$ and the bulk region $z>d_s$, where $d_s$ denotes the thickness of the surface region. Macroscopically, on a scale $\Lambda \gg d_s$ where MBCs make sense~\cite{landau}, the surface region appears extremely thin and is traditionally treated as a geometrical separation between the vacuum and the bulk region. Physically, this region, however thin, is where physical quantities undergo rapid yet regular variations~\cite{feynman}. 

To describe the dielectric responses, we look at the electric polarization $e^{i(\mathbf{k}\cdot\mathbf{r} -
\omega t)}\mathbf{p}(z)$ induced in the system by an electric field $e^{i(\mathbf{k}\cdot\mathbf{r} -
\omega t)}\mathbf{E}(z)$ present therein, where translational symmetry along the surface has been assumed, $\mathbf{r} = (x,y)$ and $\mathbf{k}$ is a planar wave vector. Needless to say, $\mathbf{p}$ vanishes in the vacuum. In the bulk region, it can be related to the electric field by a susceptibility function $S(z,z')$, i.e. $p_\mu(z>d_s) = P_\mu(z) = \sum_\nu \int^\infty_0 dz' ~ S_{\mu\nu}(z,z') E_\nu(z')$, where $\mu,\nu=x,y,z$ label the components of the fields and we have suppressed the possible dependence of $S$ on $\omega$ and $\mathbf{k}$ to simplify the notation. Note that the atomistic environment in the bulk region is supposedly indistinguishable from that in an infinite medium. Hence $S(z,z')$ must be governed by the same set of differential equations as the susceptibility function for the infinite medium. In general, $S(z,z')$ contains some parameters, which inevitably appear in the general solution to these equations, see what follows. These parameters characterize surface scattering effects and cannot be determined by the equations alone.  
 
To determine the polarization in the surface region, the obvious but often impractical option is to solve the full dynamical equations in this region, which requires the microscopic details of a surface that are unknown in reality. Fortunately, the macroscopic limit can be determined without such details. To show this, we can introduce some phenomenological functions $w_\mu(z)$ gratifying that $p_\mu(z) = w_{\mu}(z) P_\mu(z)$. By definition, $w_{\mu}(z)$ smoothly evolves from zero to unity as $z$ travels across the surface region from the vacuum into the bulk region. The exact form of $w_\mu$ depends on the microscopic details of a surface. Nevertheless, on a macroscopic scale where the surface region appears infinitely thin (i.e. $d_s/\Lambda \rightarrow 0_+$), the microscopic variations become irrelevant after standard coarse-graining~\cite{landau,feynman} and $w_{\mu}(z)$ degenerates into the Heaviside step function $\theta(z)$, i.e. $\theta(z>0) = 1$ and $\theta(z\leq0) = 0$. In this way, $\mathbf{p}(z)$ is fixed also in the surface region and hence determined throughout the system. Generalization of the reasoning to other geometries such as a slab is straightforward.  

\textit{Interaction with light}. Omitting the factor $e^{i(\mathbf{k}\cdot\mathbf{r} - \omega t)}$, the polarization charge density is obtained as $\rho(z) = - \nabla\cdot\mathbf{p}(z)$, where $\nabla = (i\mathbf{k},\partial_z)$, and the current density as $\mathbf{j}(z) = -i\omega \mathbf{p}(z)$. For isotropic materials, for which $S_{\mu\nu}(z,z') = \delta_{\mu\nu} S(z,z')$ with $\delta_{\mu\nu}$ being the Kronecker symbol, $\mathbf{P}(z)$ aligns with $\mathbf{E}(z)$. Under normal incidence ($\mathbf{k} = 0$), we may take $\mathbf{E}(z) = (E(z),0,0)$ and $\mathbf{p}(z>0) = (P(z),0,0)$. In such case the difference between $\mathbf{p}$ and $\mathbf{P}$ is immaterial since $P_z=0$. Substituting $\rho$ and $\mathbf{j}$ in Maxwell's equations gives~\cite{si}
\begin{eqnarray}
&~& E(z) = E_{in}(z) - 4\pi k^2_0\int  dz'~G(z-z') P(z'), \label{15} \\
&~& P(z) = \int dz' ~S(z,z') E(z'), \label{P}
\end{eqnarray}
where the integrals are carried out over the medium only, the Green's function $G(z) = \frac{1}{2ik_0} e^{ik_0\abs{z-z'}}$ generates ``out-going" waves and $E_{in}(z) = e^{ik_0z}$ for incident radiation source located outside the medium (e.g. in S$_1$ and S$_3$) but differs otherwise (e.g.~in S$_2$), see below. In Eq.~(\ref{P}), $z$ is confined to the medium. One may show that~\cite{si}, for non-dispersive SIM, for which $S(z,z') = S\delta(z-z')$ with $\delta(z-z')$ being the Dirac function, Eq.~(\ref{15}) reproduces the textbook result $E_r = (\sqrt{\epsilon}-1)/(\sqrt{\epsilon}+1)$ with $\epsilon = 1 + 4\pi S$ being the dielectric constant. 

\textit{Excitons by the Lorentz oscillator model}. Here we write $S(z,z') = S_b\delta(z-z') + \tilde{S}(z,z')$, where $S_b = (\epsilon_b-1)/4\pi$ represents the background response and $\tilde{S}$ accounts for the excitonic response. As aforementioned, the dynamical equation governing $\tilde{S}(z,z')$ is the same as that for an infinite medium, which can be established from the second part of Eq.~(\ref{2}) as
\begin{equation}
\left(\partial^2_z + q^2\right)\tilde{S}(z,z') = - \frac{Q^2}{4q}\delta(z-z'). \label{dye}
\end{equation} 
The solution suitable for a SIM (i.e. S$_1$ and S$_2$) must vanish at infinity. In general, it can be written as 
\begin{equation}
\tilde{S}(z,z') = S_{\infty} \left(e^{iq\abs{z-z'}} + \mathcal{R}(z') e^{iq(z+z')}\right), \label{s12}
\end{equation}
where $S_{\infty} = \frac{iQ^2}{8\pi q}$ and $\mathcal{R}(z)$ is the surface characteristic quantity, which we shall call the \textit{surface~response~function} (SRF). The first term in Eq.~(\ref{s12}) describes the out-going waves generated by an electric field localized at $z'$ and is actually the inverse Fourier transform of the second part of Eq.~(\ref{2}), which gives the response of an infinite medium. The second term describes polarization waves reflected from the surface. It represents, unlike any ABCs, an essentially non-local effect. In the widely used dielectric approximation~\cite{wolf1971,sein1972}, only the first term of Eq.~(\ref{s12}) is included. 

The function $\mathcal{R}(z)$ describes surface scattering effects and serves as a fingerprint for distinguishing one surface from another. It cannot be determined in a macroscopic theory, but can be extracted from a microscopic surface model~\cite{si} or, as shown below, from a measured optical response. 

\textit{Connection with ABCs in SIM}. The excitonic part of the total polarization is given by $\tilde{P}(z) = \int^\infty_0 dz' \tilde{S}(z,z') E(z')$. Using Eq.~(\ref{s12}), one may show that 
\begin{eqnarray}
\begin{pmatrix}
iq\tilde{P}(0) \\
\tilde{P}'(0)
\end{pmatrix}
= iq S_\infty 
\begin{pmatrix}
R + 1 \\
R - 1
\end{pmatrix} 
\int^\infty_0 dz e^{iqz} E(z), \label{ac}
\end{eqnarray}
where $R$ is an average of $\mathcal{R}(z)$, given by
\begin{equation}
\int^\infty_0 dz e^{iqz} E(z) \mathcal{R}(z) = R \int^\infty_0 dz e^{iqz} E(z). \label{defR}
\end{equation}
Note that $R$ depends on both $\mathcal{R}(z)$ and $E(z)$, the latter being specific to the way the system is optically excited. Equation~(\ref{ac}) implies that
\begin{equation}
\tilde{P}'(0) = \kappa \tilde{P}(0), \quad \kappa = iq\frac{R-1}{R+1}, \label{abc}
\end{equation}
which has apparently the same form as a general ABC supposedly characterizing a surface~\cite{kiselev1975,Johnson1976,egor2002}. A big caveat here is that $R$ (and $\kappa$) can not be interpreted as a surface characteristic (material parameter). Indeed, $R$ not just varies from one surface to another but, depending on the details of the way the system is excited, could take on different values even for the same surface. In what follows, we show this for both the SIM and slab geometry. It shall be seen that Eq.~(\ref{abc}) (and also $\tilde{S}$) needs to be modified for a slab merely due to the existence of two surfaces, which allow waves to travel back and forth. This again underlines that $R$ is not a surface property.  

\textit{Results for S$_1$ and S$_2$}. We begin with S$_1$, where the incident light impinges on the surface from outside, as depicted in Fig.~\ref{fig:0} (a). The solution to Eqs.~(\ref{15}) and (\ref{P}) is obtained from the ansatz that for $z$ lying in the medium, $P(z) = \sum_j P_j e^{ik_j z}$ and $E(z) = \sum_j E_j e^{ik_j z}$, where Im$(k_j)\geq0$. Substituting this in the equations leads to
\begin{equation}
E_j = \frac{4\pi k^2_0}{k^2_j - k^2_0}P_j, \quad P_j = S_j E_j, \quad S_j = \frac{\epsilon(k_j) - 1}{4\pi}, \label{kalpha}
\end{equation}
where $\epsilon(k)$ is given by Eq.~(\ref{2}), and 
\begin{align}
\sum_{j = 1,2} E_j \left(\frac{1}{k_j - q} + \frac{\mathcal{R}_j(q)}{k_j + q}\right) = 0, \label{bda} \\
\sum_{j = 1,2} E_j (k_j + k_0) = 2k_0, \label{bdb}
\end{align}
with $\mathcal{R}_j(q) = -i(q+k_j) \int^\infty_0 dz'~ \mathcal{R}(z') e^{i(q+k_j)z'}$. Equation~(\ref{kalpha}) does not explicitly involve $\mathcal{R}_j$ and has the same form as for an infinite medium, in consistency with Ewald-Oseen extinction theorem~\cite{wolf1972}. Hence, $k_j$ are the roots of Eq.~(\ref{3}) as proposed by Pekar. Equations (\ref{bda}) and (\ref{bdb}) uniquely fix the amplitudes $E_j$. The former alone takes care of boundary effects while the latter can be shown equivalent to the MBCs.   

\begin{figure}
\begin{center}
\includegraphics[width=0.45\textwidth]{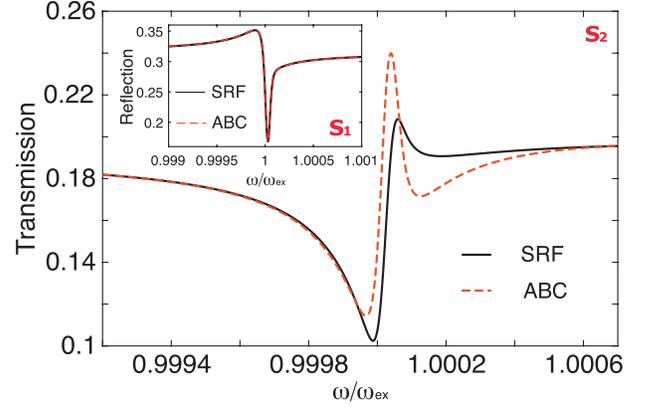}
\end{center}
\caption{Transmission $\abs{E_t}^2 = \abs{E(0)}^2$ of light into vacuum for S$_2$ with $\mathcal{R}(z) = - e^{-sz}$, where $sQ^{-1} = 0.05 + 0.32i$ and $z_sQ = 4.75$, by SRF theory [i.e. Eqs.~(\ref{ins1}) - (\ref{ins})] and ABC [i.e. Eqs.~(\ref{ins1}) - (\ref{w}),  (\ref{abc}) and (\ref{r})]. Inset: Reflection $\abs{E_r}^2$ for S$_1$, with $E_r = E(0)-1$, by SRF theory [Eqs.~(\ref{bda}) and (\ref{bdb})] and ABC [Eqs.~(\ref{abc}) and (\ref{r})]. The same $\mathcal{R}(z)$ is used. \label{fig:1}}
\end{figure}

Inserting the ansatz into Eq.~(\ref{defR}) shows that $\sum_j \frac{E_j \mathcal{R}_j(q)}{k_j+q} = R \sum_j \frac{E_j}{k_j + q}$. With this, Eq.~(\ref{bda}) can be rewritten as
\begin{equation}
\sum_{j=1,2} E_j \left(\frac{1}{k_j - q} + \frac{R}{k_j + q}\right) = 0. \label{R}
\end{equation}
which is equivalent to Eq.~(\ref{abc}). The equivalence between Eqs.~(\ref{bda}) and (\ref{R}) fixes $R$, yielding
\begin{equation}
R = \frac{(k_2+q)(k_1-q)\mathcal{R}_1 - (k_2-q)(k_1+q)\mathcal{R}_2}{(\mathcal{R}_2-\mathcal{R}_1)(k_2-q)(k_1-q) + 2q(k_1-k_2)}. \label{r} 
\end{equation}
The optical responses of the system are obtained by solving Eqs.~(\ref{kalpha}), (\ref{bda}) and (\ref{bdb}). An example is shown in the inset of Fig.~\ref{fig:1}, where the reflection $\abs{E_r}^2$ is plotted for $\mathcal{R}(z) = - e^{-sz}$ with Re$(s)>0$. We see that the results by Eq.~(\ref{bda}) are the same as by Eq.~(\ref{R}) for any $s$ with $R$ given by Eq.~(\ref{r}).    

The above analysis shows that S$_1$ is not sensitive to the whole profile of $\mathcal{R}(z)$ but only probes the average $R_\infty$. To experimentally extract $\mathcal{R}(z)$, it is necessary to analyze setups which detect the full $\mathcal{R}(z)$. S$_2$ and S$_3$ each suffice for this purpose. In S$_2$ light is incident from a source located at $z=z_s>0$ inside the medium, see Fig.~\ref{fig:0} (b), for which we have~\cite{si} $E_{in} = e^{ik_0|z-z_s|}$. To solve Eqs.~(\ref{15}) and (\ref{P}), the ansatz is made that $P(0<z<z_s) = \sum_\alpha P_\alpha e^{ik_\alpha z}$, $P(z>z_s) = \sum_j P_j e^{ik_j z}$, and $E(0<z<z_s) = \sum_\alpha E_\alpha e^{ik_\alpha z}$ as well as $E(z>z_s) = \sum_j E_j e^{ik_j z}$. There are no restrictions on $k_\alpha$ but Im$(k_j)\geq0$. Substituting them in the equations reveals that $E_{\alpha/j} = \frac{4\pi k^2_0}{k^2_{\alpha/j} - k^2_0} P_{\alpha/j}$, $P_{\alpha/j} = S_{\alpha/j} E_{\alpha/j}$ with $S_{\alpha/j} = \frac{\epsilon(k_{\alpha/j})-1}{4\pi}$, which have the same form as Eq.~(\ref{kalpha}), again confirming the extinction theorem. Both $k_\alpha$ and $k_j$ obey Eq.~(\ref{3}), yielding four $\alpha$-modes with amplitudes $E_\alpha$ and two $j$-modes with amplitudes $E_j$, which are determined by
\begin{eqnarray}
&~& \sum_\alpha E_\alpha (k_\alpha + k_0) = 0, \label{ins1} \\
&~& \sum_je^{ik_jz_s}E_j(k_j\pm k_0) - \sum_\alpha e^{ik_\alpha z_s}E_\alpha(k_\alpha\pm k_0) = 2k_0, \\
&~& \sum_je^{ik_jz_s}\frac{E_j}{k_j\pm q} - \sum_\alpha e^{ik_\alpha z_s}\frac{E_\alpha}{k_\alpha\pm q} = 0, \label{w} \\
&~& \sum_\alpha E_\alpha \left(\frac{1}{k_\alpha - q} + \frac{\mathcal{R}_\alpha(z_s,q)}{k_\alpha + q}\right) + \sum_j E_j \frac{\mathcal{R}_j(z_s,q)}{k_j + q} = 0, \label{ins}
\end{eqnarray}
where $\mathcal{R}_\alpha(z_s,q) = -i(k_\alpha + q) \int^{z_s}_0 dz'~ \mathcal{R}(z') e^{i(k_\alpha + q)z'}$ and $\mathcal{R}_j(z_s,q) = -i(k_j + q) \int^\infty_{z_s} dz'~ \mathcal{R}(z') e^{i(k_j + q)z'}$. These equations can be compactly rewritten in matrix form, $M(z_s) E = \psi$, where $M(z_s)$ is a 6-by-6 matrix with elements provided in Ref.~\cite{si}, and $E$ is a column vector with elements $E_\alpha$ and $E_j$ while $\psi$ is also a column vector with all elements vanishing except two, which both are $2k_0$. 

Inserting the ansatz for S$_2$ into Eq.~(\ref{defR}) and using Eq.~(\ref{w}) reveal that $\sum_\alpha E_\alpha \frac{\mathcal{R}_\alpha(z_s,q)}{k_\alpha + q} + \sum_j E_j \frac{\mathcal{R}_j(z_s,q)}{k_j + q} = \sum_\alpha E_\alpha \frac{R}{k_\alpha + q}$. With this, Eq.~(\ref{ins}) can again be transformed into Eq.~(\ref{R}) but with $j$ replaced by $\alpha$. Hence, the ABC (\ref{abc}) remains valid. Nevertheless, there is a crucial difference resting with the fact that here $R$ generally is not equal to the average given in Eq.~(\ref{r}). Instead, it takes on a completely different value~\cite{si} and varies with $z_s$. Exemplifying this discrepancy, we have calculated the transmission (of light into vacuum) as $\abs{E_t}^2 = \abs{E(0)}^2$ for $\mathcal{R}(z) = - e^{-sz}$ first by the proper theory [i.e. Eqs.~(\ref{ins1}) - (\ref{ins})] and then by the ABC [Eqs.~(\ref{ins1}) - (\ref{w}) and  (\ref{R})] with $R$ given by Eq.~(\ref{r}). The results are displayed in Fig.~\ref{fig:1} and they are clearly different. This shows that, experimentalists cannot use the value of $R$, which they have painstakingly measured with S$_1$, to make predictions regarding the outcome for S$_2$ even though the same system is experimented with. The situation is made worse by the dependence of $R$ on $z_s$, which nullifies their effort to predict what would happen if the radiation source is displaced. 

It should be noted that, only in the special case of a constant $\mathcal{R}(z) = R_\infty$, the value of $R$ is the same for both S$_1$ and S$_2$ and equal to $R_\infty$. 

\begin{figure}
\begin{center}
\includegraphics[width=0.49\textwidth]{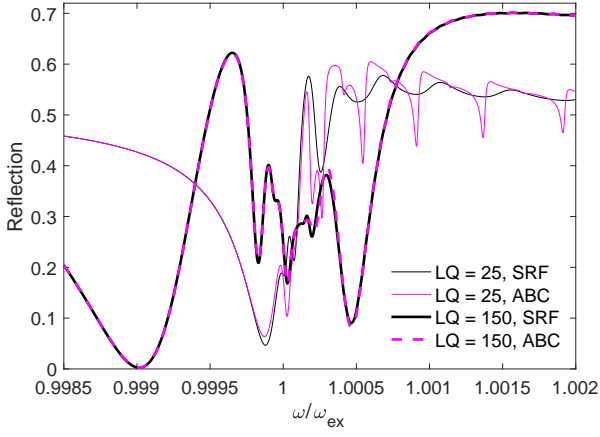}
\end{center}
\caption{Slab reflection $\abs{E_r}^2$ for the same $\mathcal{R}(z)$ as in Fig.~\ref{fig:1}. SRF Theory: Eqs.~(\ref{slab0}) - (\ref{slab2}). ABC: Eqs.~(\ref{slab0}) and (\ref{slababc}) with $\kappa$ given by Eq.~(\ref{abc}) and $R$ by Eq.~(\ref{r}).\label{fig:4}}
\end{figure}   

\textit{Slab systems: S$_3$}. A slab has two surfaces, which we assume are located at $z=0$ and $z=L>0$, respectively, as shown in Fig.~\ref{fig:0} (c). The exciton response function $\tilde{S}(z,z')$ now contains an extra contribution that represents the reflected polarization waves from the surface at $z=L$, 
\begin{equation}
\tilde{S}(z,z') = S_{\infty} \left(e^{iq\abs{z-z'}} + \underline{\mathcal{R}}(z') e^{iq(z+z')} + \underline{\mathcal{R}}(\bar{z}')e^{iq(\bar{z}+\bar{z}')}\right), \label{slab}
\end{equation}
where $\bar{z} = L-z$ and $\bar{z}' = L - z'$, and the slab is assumed symmetric under reflections about its mid-plane $z=L/2$. It can be shown that~\cite{si} 
\begin{equation}
\underline{\mathcal{R}}(z) = \frac{\mathcal{R}(z)+e^{2iq(L-z)}\mathcal{R}(L)\mathcal{R}(L-z)}{1-\mathcal{R}^2(L)e^{2iqL}}, 
\end{equation}
which can never be a constant except for identically vanishing $\mathcal{R}(z)$, accounting for multiple reflections of polarization waves by the two surfaces. In the semi-infinite limit $\underline{\mathcal{R}}(z)$ tends to $\mathcal{R}(z)$. Clearly, $\mathcal{R}(z)$ is a pure characteristic of a single surface whereas $\underline{\mathcal{R}}(z)$ represents a cumulative effect of both.  

Equations (\ref{15}) and (\ref{P}) are solved by the same ansatz as for S$_1$ but with no restrictions on wave numbers $k_\alpha$ here. In agreement with the extinction theorem~\cite{wolf1972}, Eq.~(\ref{kalpha}) remains valid so $k_\alpha$ are now all four roots of Eq.~(\ref{3}). Equations~(\ref{bda}) and (\ref{bdb}) are now augmented as follows,
\begin{eqnarray}
&~& \sum_\alpha E_\alpha(k_\alpha + k_0) = 2k_0, \quad \sum_\alpha e^{ik_\alpha L}E_\alpha (k_\alpha - k_0) = 0, \label{slab0} \\
&~& \sum_\alpha E_\alpha \left(\frac{1}{k_\alpha - q} + \frac{\mathcal{R}^+_\alpha(q,L)}{k_\alpha + q}\right) = 0, \label{slab1} \\
&~& \sum_\alpha e^{ik_\alpha L}E_\alpha \left(\frac{\mathcal{R}^-_\alpha(q,L)}{k_\alpha - q} + \frac{1}{k_\alpha + q}\right) = 0. \label{slab2} 
\end{eqnarray}
Here $\mathcal{R}^{\pm}_\alpha(q,L) = -i(q\pm k_\alpha) \int^L_0 dz'~\underline{\mathcal{R}}(z') e^{i(q\pm k_\alpha)z'}$ depends on both $q$ and $L$. It can be shown~\cite{si} that, in the semi-infinite limit $L\rightarrow\infty$, $E_\alpha$ vanishes for modes with Im$(k_\alpha)<0$ and the results for S$_1$ are restored. 

Mistaking that ABCs represent surface characteristics, one might simply apply the ABC (\ref{abc}) to the slab surfaces separately by imposing that~\cite{egor2002}  
\begin{equation}
\tilde{P}'(0) = \kappa \tilde{P}(0), \quad \tilde{P}'(L) = -\kappa \tilde{P}(L). \label{slababc}
\end{equation}
However, these conditions can be easily and rigorously shown incompatible with Eqs.~(\ref{slab1}) and (\ref{slab2}). As an illustration, in Fig.~\ref{fig:4} we display the reflection $|E_r|^2$ calculated for the same $\mathcal{R}(z)$ as in Fig.~\ref{fig:1}. The results by the SRF theory [i.e. Eqs.~(\ref{slab0}) - (\ref{slab2})] are obviously different from those by the ABC [i.e. Eqs.~(\ref{slab0}) and (\ref{slababc})] even for a not so thick slab. Actually, for moderately thin slabs ABC produces unphysical results~\cite{si}. For thick slabs our theory and the ABC produce close results thanks to effective decoupling of the surfaces. For very thick slabs, the results for S$_1$ are recovered~\cite{si}. 

Notwithstanding, a generalization of Eq.~(\ref{slababc}) can be shown compatible with the SRF theory. To see this, we use Eq.~(\ref{slab}) to obtain the exciton polarization and find
\begin{equation}
\begin{pmatrix}
iq\tilde{P}(0) \\
iq\tilde{P}(L) \\
\tilde{P}'(0) \\
\tilde{P'}(L)
\end{pmatrix}
= iqS_\infty
\begin{pmatrix}
1+R_0, & e^{iqL}R_L \\
-R_0e^{iqL}, & (1+R_L) \\
R_0 - 1, & -R_L e^{iqL} \\
R_0e^{iqL}, & (1-R_L)
\end{pmatrix}
\begin{pmatrix}
\int^L_0 dz e^{iqz} E(z) \\
\int^L_0 dz e^{iq(L-z)} E(z)
\end{pmatrix}, \label{gelabc}
\end{equation}
which is just the analogue of Eq.~(\ref{ac}). Here the quantities $R_{0/L}$ are counterparts of $R$, defined as
\begin{subequations}
\begin{align}
\int^L_0 dz e^{iqz} \underline{\mathcal{R}}(z) E(z) = R_0 \int^L_0 dz e^{iqz} E(z), \\
\int^L_0 dz e^{iq(L-z)} \underline{\mathcal{R}}(L-z) E(z) = R_L \int^L_0 dz e^{iq(L-z)} E(z). 
\end{align}
\label{r0l}
\end{subequations}
Eliminating the integrals from Eq.~(\ref{r0l}) yields
\begin{equation}
\tilde{P}'(0) = \kappa_0 \tilde{P}(0) + \gamma \tilde{P}(L), \quad \tilde{P}'(L) = - \kappa_L \tilde{P}(L) - \gamma \tilde{P}(0), \label{gelabc}
\end{equation}
which, being equivalent to Eqs.~(\ref{slab1}) and (\ref{slab2}), generalizes Eq.~(\ref{slababc}). In Eq.~(\ref{gelabc}),
\begin{equation}
\begin{pmatrix}
\kappa_0 \\
\kappa_L \\
\gamma
\end{pmatrix}
= \frac{iq}{D}
\begin{pmatrix}
(1-R_0)(1+R_L) - R_0 R_Le^{2iqL} \\
(1-R_L)(1+R_0) - R_0 R_Le^{2iqL} \\
2R_0R_L e^{iqL}
\end{pmatrix}
\end{equation}
with $D = R_0R_Le^{2iqL}-(1+R_0)(1+R_L)$. In general $\kappa_0 \neq \kappa_L$ except in the semi-infinite limit. This asymmetry stems from the setup, not the slab geometry itself, which highlights again that these parameters are not surface characteristics. Like $R(z_s)$ varying with $z_s$, $R_0$ and $R_L$ vary with $L$. 

\textit{Experimental relevance of $\mathcal{R}(z)$}. Finally, we discuss how to experimentally extract the SRF $\mathcal{R}(z)$. This is impossible with S$_1$ but possible with S$_2$ and S$_3$. 

We take S$_2$ for instance. Let us assume that $\mathcal{R}(z)$ is a decaying function of $z$ so that it is vanishingly small for $z$ beyond certain value $z_{max}$, i.e. $\mathcal{R}(z\geq z_{max}) \approx 0$. For $z<z_{max}$, we may represent it as a truncated Fourier series, namely $\mathcal{R}(z<z_{max}) \approx \sum^N_{n=0} c_n \cos\left(\frac{n\pi}{z_{max}}z\right)$, where $N$ determines the spatial resolution. Note that $c_N$ can be obtained from other coefficients by the constraint that $\mathcal{R}(z_{max}) = 0$ and is hence not a free parameter. As such, $\mathcal{R}_\alpha(z_s,q)$ and $\mathcal{R}_j(z_s,q)$ can be written as linear combinations of the Fourier coefficients, i.e. $\mathcal{R}_{\alpha/j}(z_s,q) = \sum^{N-1}_{n=0} \lambda_{\alpha/j,n}(z_s,z_{max}) c_n$. The expressions of $\lambda_{\alpha/j,n}$ are not quoted here. With this the matrix $M(z_s)$ becomes a function of $c=(c_0,c_1,...,c_{N-1})$ and $z_{max}$. 

To extract $c$ and $z_{max}$ experimentally, we look at the amplitude of light transmitted into the vacuum [see Fig.~\ref{fig:0} (b)] for $(N+1)$ different values of $z_s$, say $z_{s,0}, z_{s,1}, ..., z_{s,N}$. The transmission amplitude for $z_{s,l}$ is denoted by $E_t(l)$, which can be experimentally measured. Our theory gives $E_t(l) = \langle V M^{-1}(z_{s,l}) \psi\rangle$, where $V =$~diag$(1,1,1,1,0,0)$ and $<...>$ sums all the elements. With measured $E_t(l)$, these relations constitute $(N+1)$ implicit equations for $c$ and $z_{max}$ and can be solved to determine $\mathcal{R}(z)$ with a spatial resolution $\sim z_{max}/N$. Solving the equations is equivalent to minimizing the following expression,
\begin{equation}
\mathcal{F}(c_0,...,c_N) = \sum^N_{l=0} \abs{E_t(l) - \langle V M^{-1}(z_{s,l}) \psi\rangle}^2. 
\end{equation}
This is a standard least squares problem and may be solved using a variety of algorithms~\cite{jorge}. 

If $\mathcal{R}(z)$ does not decay fast, a better way of sampling is perhaps to expand it as a sum of exponentials rather than Fourier series, i.e. $\mathcal{R}(z) = \sum^N_{n=1} c_n e^{iQ_nz}$, where both $c_n$ and $Q_n$ need to be determined. This requires measuring $2N$ instead of $N$ transmission amplitudes, one for each value of $z_s$. 

A main difficulty of extracting the SRF with S$_2$ rests with the implantation and control of the interior radiation source. Using S$_3$ avoids this difficulty but requires preparing a multitude of samples~\cite{si}. In the future, we shall explore more practical methods based on oblique incidence and ellipsometry as well as interference effects.    

\textit{Conclusions}. A macroscopic theory has been presented for dealing with the optical responses of bounded dispersive medium without invoking any ABCs. ABCs are shown bearing no direct relation to physically meaningful parameters of the system but can be deduced, if needed, from the surface response function $\mathcal{R}(z)$ introduced in this work. This function is an intrinsic property of a surface, which, unlike any ABCs, reflects the generally non-local contribution of the surface to the optical response of the system. Experimental procedures have been proposed for mapping out this function. 

Our results call for a reappraisal of innumerable experiments that have been interpreted on the basis of ABCs. An interesting direction for future research may be to apply the theory to both ordinary and topological metamaterials, which have recently attracted much attention due to their strongly dispersive and anisotropic electrodynamic responses~\cite{pendry1996, silve2003, silve2009, zhang2018, zhang2020, zhang}.

\end{document}